\documentclass[a4paper,11pt]{article}
\usepackage{jheppub} 

\usepackage[T1]{fontenc} 
\usepackage{graphicx}
\graphicspath{ {figures/} }
\usepackage{amsmath}
\usepackage{bm}
\usepackage{subfigure}
\usepackage{amssymb}
\usepackage{mathrsfs}
\usepackage{braket}
\usepackage{orcidlink}
\usepackage{comment}

\newcommand{\comments}[1]{}
\usepackage{slashed}
\usepackage{epsfig,psfrag,rotating,soul}
\usepackage{rotfloat}
\usepackage{empheq}  

\usepackage[utf8]{inputenc}  
\usepackage{tensor}
\usepackage{hyperref}
\usepackage{microtype}

\title{\boldmath Regular Reissner--Nordström--de Sitter Black Holes: Thin-Shell Matching and Modified Density Profiles}

\author{Reza Saadati\orcidlink{0000-0001-5657-6023}$^{1,2}$, Yu Shi\orcidlink{0000-0002-2910-0684}$^{1,2}$ and Vitalii Vertogradov\orcidlink{0000-0002-5096-7696}$^{1,2,3,4,5}$}

\affiliation{$^1$Wilczek Quantum Center, Shanghai Institute for Advanced Studies, Shanghai 201315, China}

\affiliation{$^2$University of Science and Technology of China, Hefei 230026, China}

\affiliation{$^3$Physics department, Herzen state Pedagogical University of Russia,
48 Moika Emb., Saint Petersburg 191186, Russia.}
\affiliation{$^4$Center for Theoretical Physics, Khazar University, 41 Mehseti Street, Baku, AZ-1096, Azerbaijan.}

\affiliation{$^5$SPB branch of SAO RAS, 65 Pulkovskoe Rd, Saint Petersburg
196140, Russia.}

\emailAdd{yu\_shi@ustc.edu.cn}

\abstract{We investigate two distinct approaches to regularizing the central singularity of the Reissner--Nordström--de Sitter geometry. In the first approach, a de Sitter interior is matched to an exact Reissner--Nordström--de Sitter exterior across a timelike dust shell. Using the Israel junction conditions, we derive the effective potential governing the shell dynamics and analyze its
static equilibrium and radial stability. We show that radial stability requires $\Lambda_->\Lambda_+$, while the additional requirement of three distinct exterior horizons, with the shell located inside the inner horizon, leads to the stronger analytical result that such stable
black-hole configurations exist if and only if $\Lambda_->3\Lambda_+$. We also derive an upper bound on the exterior mass parameter and recover the Reissner--Nordström limit as $\Lambda_+\to0$. Although the singular center is replaced by a regular core, the stationary exterior geometry remains unchanged, demonstrating that exterior geometric observations alone
cannot determine whether the black-hole interior is singular or regular. In the second approach, we modify the divergent energy-density profile and reconstruct a smooth geometry with a de Sitter core while preserving the Reissner--Nordström--de Sitter asymptotics. Regularity fixes the smoothing scale to $\beta=Q^2/(2m)$, leaving no independent regularization parameter. The two constructions thus provide distinct regularization mechanisms, rather than equivalent descriptions of the same spacetime.}

\begin{document}
\maketitle
\flushbottom

\section{Introduction}
The singularity problem remains one of the central unresolved issues in the classical theory of black holes. After the observations of the Event Horizon Telescope~\cite{bib:eht1,bib:eht2}, black holes acquired the status of  observed astrophysical objects. Nevertheless, the standard classical solutions describing black holes contain singular regions where the description in terms of general relativity breaks down. This motivates two broad directions: either to search for new solutions of the Einstein equations describing nonsingular, or regular, black holes, or to modify the underlying gravitational theory itself.

According to the Penrose singularity theorem~\cite{bib:penrose}, singularity formation is a generic consequence of gravitational collapse once the appropriate focusing condition is satisfied. In the present effective-matter context this requirement may be formulated in terms of the strong energy condition. Moreover, several studies have shown that singularity formation may become observationally relevant if the formation of apparent horizons is delayed~\cite{bib:joshi,bib:joshi_review,bib:dei2022epjc,bib:inhomo,bib:inhomo2}. Therefore, regular black holes are not merely formal alternatives to singular geometries; they also provide a useful framework for testing how high-density matter may alter the final stages of gravitational collapse.

The first regular black hole model was proposed phenomenologically by Bardeen~\cite{bib:bardeen}. Since this solution was not originally derived from a specified matter source, the physical origin of the stress-energy tensor supporting it remained unclear for a long time. It was later realized that the Bardeen geometry can be interpreted as a solution sourced by nonlinear electrodynamics~\cite{bib:garcia}. However, this interpretation has a  limitation: if nonlinear electrodynamics is used as the source of a regular black hole, purely electric regular solutions are strongly constrained, and the regularizing charge is typically interpreted as a magnetic monopole~\cite{bib:bronnikov}. This feature may lead to additional difficulties, including possible singular behavior in charged processes of Penrose type~\cite{bib:angel2025podu}.

An alternative physical idea goes back to Gliner~\cite{bib:gliner} and Sakharov~\cite{bib:sakharov}. They suggested that, at sufficiently high densities, matter may undergo a transition to a vacuum-like state described by the equation of state $P=-\rho$.
Such a state generates a de Sitter core and can prevent the formation of a central curvature singularity. This mechanism has been developed in many models of regular black holes with de Sitter interiors~\cite{bib:dym,bib:dym2,bib:hay,bib:khlopov1,bib:khlopov2,bib:khlopov3,bib:khlopov4,bib:khlopov5,bib:ali2025cqg}; see also the reviews~\cite{bib:ansoldy,bib:regular}.

The existence of a de Sitter core is closely connected with the violation of the strong energy condition. Indeed, if the strong energy condition remains valid throughout the collapsing matter distribution, the singularity theorem implies that a regular center cannot be obtained. Hence, any regular black hole supported within general relativity must violate the strong energy condition in the central region~\cite{bib:zaslav}. Since ordinary stellar matter does not possess such vacuum-like properties at low density, the violation of the strong energy condition should be interpreted as an effective high-density phenomenon emerging during the late stages of collapse~\cite{bib:vertogradov2025podu,bib:ali2025jcap,bib:vertogradov2025epjc}.

In this work, we focus on the regularization of the Reissner--Nordström--de Sitter geometry. In its standard form this spacetime contains a central singularity caused by the divergent electromagnetic contribution to the energy density. Our approach to this problem is twofold. First, we use the Israel junction formalism~\cite{Barrabes:1991ng, Israel:1966rt} to replace the singular central region of the black hole with a regular de Sitter core separated from the exterior geometry by a thin dust shell.  This construction is closely related to a long line of thin-shell models in which singular black hole interiors are replaced by regular cores. In particular, de Sitter interiors joined to exterior black hole geometries have been extensively studied in the context of gravastars and thin shell compact objects~\cite{Mazur:2004fk,Mottola:2023jxl,Visser:2003ge,MartinMoruno:2011rm}. Charged versions of such configurations are especially relevant to the present work. Regular charged black holes with a de Sitter core and a Reissner--Nordström exterior were constructed~\cite{Lemos:2011dq}, while its stability was analyzed~\cite{Uchikata:2012zs}. Our analysis differs from these works by considering a Reissner--Nordström--de Sitter exterior and by showing explicitly that stable dust shell configurations require different values of the cosmological constant on the two sides of the shell. Since the regular core is hidden behind the event horizon, the thin-shell construction does not produce directly observable deviations from the stationary exterior Reissner--Nordström--de Sitter geometry. Nevertheless, it demonstrates that an identical exterior geometry can be compatible with a regular interior, while allowing us to investigate how the difference between the interior and exterior vacuum energies constrains the existence and radial stability
of such configurations.

In the second approach, we show that the singular behavior can be removed by modifying the density profile in the high-curvature region while preserving the Reissner--Nordström--de Sitter asymptotics at large radii. The modification introduces a finite central energy density and drives the matter sector toward a de Sitter-type equation of state near the origin. In this way, the central singularity is replaced by a regular vacuum-like core.

A key point of the construction is that the regularization is not introduced directly at the level of the metric function. Instead, we deform the matter density and then reconstruct the corresponding mass function from the Einstein equations. This makes the physical origin of the regular geometry more transparent. The resulting effective matter distribution reproduces the usual Reissner--Nordström--de Sitter equation of state in the limit where the regularization parameter is removed, whereas at small radii it acquires an additional nonlinear contribution. This contribution can be written as a generalized polytropic-type equation of state and is responsible for the transition from the singular electromagnetic profile to a nonsingular de Sitter core.

\section{Junction formalism}
We construct regular black hole configurations by matching two static, spherically symmetric spacetimes across a timelike hypersurface using the Israel junction conditions~\cite{Barrabes:1991ng,Israel:1966rt}. Let $\mathcal{V}^-$ and $\mathcal{V}^+$ denote the interior and exterior regions with the line elements
\begin{equation}\label{metric_pm}
    ds_\pm^2 = - f_\pm(r)\, dt_\pm^2 + \frac{dr^2}{f_\pm(r)} + r^2 d\Omega^2 ,
\end{equation}
where $d\Omega^2$ is the metric on the unit two-sphere. The hypersurface $\Sigma$ is defined by $r=R(\tau)$, where $\tau$ is the proper time of an observer co-moving with the shell. The induced metric on $\Sigma$, expressed in terms of intrinsic coordinates $\left(\tau,\theta,\phi\right)$, must be continuous, so 
\begin{equation}
    [h_{ab}] = 0,
\end{equation}
which yields
\begin{equation}\label{induced_metric}
    ds_\Sigma^2 = - d\tau^2 + R(\tau)^2 d\Omega^2 .
\end{equation}
The normalization of the shell four-velocity implies
\begin{equation}\label{velocity_norm}
    - f_\pm(R)\,\dot t_\pm^2 + \frac{\dot R^2}{f_\pm(R)} = -1 ,
\end{equation}
where a dot denotes differentiation with respect to $\tau$. Solving for $\dot{t}_{\pm}$ gives
\begin{equation}\label{tdot}
    \dot t_\pm = \frac{\sqrt{\dot R^2 + f_\pm(R)}}{f_\pm(R)} .
\end{equation}

The unit normal vectors to $\Sigma$ are chosen to point from $\mathcal{V}^-$ to $\mathcal{V}^+$, corresponding to the direction of increasing radial coordinate on both sides of the shell. They are given by
\begin{equation}
    n_\mu^\pm=(-\dot R,\dot t_\pm,0,0),
\end{equation}
and satisfy $n_\mu n^\mu=1$ and $n_\mu u^\mu=0$.
Adopting the convention
$K_{ab}^{\pm}=e_a^{\mu}e_b^{\nu}\nabla_\mu n_\nu^{\pm}$,
the nonzero components of the extrinsic curvature are
\begin{equation}\label{K}
    {K^{\pm}}\indices{^\tau_\tau}=\frac{\ddot R+f_\pm'(R)/2}{\beta_\pm},
    \qquad    {K^{\pm}}\indices{^\theta_\theta}={K^{\pm}}\indices{^\phi_\phi} =\frac{\beta_\pm}{R},
\end{equation}
where
\begin{equation}\label{beta_def}
    \beta_\pm=\sqrt{\dot R^2+f_\pm(R)}.
\end{equation}
The Israel junction conditions relate the jump of the extrinsic curvature to the surface stress tensor $S_{ab}$,
\begin{equation}\label{israel_general}
    [K_{ab}] = -8\pi \left(S_{ab} - \frac{1}{2} h_{ab} S \right) .
\end{equation}
We consider a pressureless thin shell,
\begin{equation}
    S_{ab} = \sigma u_a u_b ,
\end{equation}
where $\sigma$ is the surface energy density and $u^a$ is the shell three-velocity. Spherical symmetry implies that the angular components of
Eq.~\eqref{israel_general} give
\begin{equation}
    [K\indices{^\theta_\theta}] =\frac{\beta_+-\beta_-}{R} =-4\pi\sigma.
\end{equation}
Consequently, the independent junction equation becomes
\begin{equation}\label{junction_sigma}
    \beta_- - \beta_+ = 4\pi R\sigma.
\end{equation}
Conservation of the surface stress-energy, $D_a S^{ab}=0$, implies
\begin{equation}
    \frac{d}{d\tau}(R^2 \sigma) = 0 ,
\end{equation}
so that the proper mass of the shell,
\begin{equation}\label{M_def}
    M = 4\pi R^2 \sigma ,
\end{equation}
is constant along the motion. In terms of $M$, the shell equation of motion becomes
\begin{equation}\label{master_eq}
    \beta_- - \beta_+ = \frac{M}{R} .
\end{equation}
Equation~\eqref{master_eq} completely determines the radial dynamics once the metric functions $f_{\pm}(r)$ are specified. Squaring Eq.~\eqref{master_eq} and rearranging yields a first-order equation of motion in the form
\begin{equation}\label{energy_form}
    \dot R^2 + V(R) = 0 ,
\end{equation}
with effective potential
\begin{equation}\label{V_general}
V(R)= \frac{4R^4f_- f_+ -\left[M^2-\left(f_++f_-\right)R^2\right]^2}{4 M^2 R^2}
\end{equation}
Static configurations correspond to radii $R_0$ satisfying
\begin{equation}
    V(R_0)=0, \qquad V'(R_0)=0
\end{equation}
while linear stability requires
\begin{equation}
    V''(R_0)>0.
\end{equation}
In the following section, we specify the interior and exterior geometries and analyze the existence and stability of stationary solutions.

\section{Regular Reissner--Nordström--de Sitter Configuration}\label{sec:Regular Reissner--Nordström--de Sitter Configuration}
We now specialize to the case where the interior geometry is de Sitter spacetime
\begin{equation}\label{fminus}
    f_-(r) = 1 - \frac{r^2}{L_-^2},
\end{equation}
while the exterior geometry is Reissner--Nordström--de Sitter,
\begin{align}\label{fplus}
    f_+ &=1-\frac{2\mathcal{M}(r)}{r}\\ \nonumber
    \mathcal{M}(r)&=m-\frac{Q^2}{2r}+\frac{r^3}{2L_+^2}
\end{align}
Here $m$ and $Q$ denote the mass parameter and electric charge of the exterior spacetime, respectively, while $L_\pm=\sqrt{3/\Lambda_\pm}$ are the curvature radii associated with the interior and exterior cosmological constants. In the asymptotically flat limit, $\Lambda_+\to 0$, the parameter $m$ coincides with the ADM mass. Substituting Eqs.~\eqref{fminus} and~\eqref{fplus} into~\eqref{energy_form}, we obtain the dynamics of the shell. In this situation, the effective potential is given by
\begin{equation}
    V(R)=1-\frac{R^2}{L_-^2}-\left[\frac{M}{2R}-\frac{1}{2M}\left(\frac{Q^2}{R}-2m+\frac{R^3}{L_-^2}-\frac{R^3}{L_+^2}\right)\right]^2.
\end{equation}
For a static timelike shell, the metric functions must satisfy $f_\pm(R_0)>0$. In the interior de Sitter geometry, this requires $R_0<L_-$. For an exterior Reissner--Nordström--de Sitter black hole with three distinct positive horizons $r_-<r_+<r_c$, the metric function is positive in the two regions $r<r_-$ and $r_+<r<r_c$. In the present construction, the shell is placed in the inner static region, $R_0<r_-$, behind the black-hole event horizon. Thus, the shell is located inside both the interior de Sitter cosmological horizon and the exterior inner horizon. Furthermore, the junction condition imposes the following constraint on the shell mass and radius:
\begin{equation}\label{Shell Mass Bound}
    M<R\sqrt{1-\frac{R^2}{L_-^2}}.
\end{equation}
This yields an upper bound for the shell mass as $M<L_-/2$. Similar inequalities have been obtained in the context of regular Kiselev black hole~\cite{Saadati:2020dje}, although the underlying matter content differs from the present model. After some algebra, the second derivative of the effective potential at the equilibrium takes the form
\begin{equation}\label{eq:RNL ddv}
	V''(R)=8-\frac{2 L_-^2}{L_-^2-R^2}-\frac{12 L_-^2 \left[L_+^2 (m-R)+L_-^2 R\right]}{L_-^2 \left(L_+^2 m+2 R^3\right)-2 L_+^2 R^3}.
\end{equation}
Linear stability requires $V''(R_0)>0$ at the static radius $R_0$. Inspection of Eq.~\eqref{eq:RNL ddv} shows that $V''(R)$ cannot be positive when $L_+<L_-$. In terms of the cosmological constants, this condition corresponds to $\Lambda_+>\Lambda_-$. Hence, a necessary condition for the existence of linearly stable configurations is
\begin{equation}
    L_+ > L_- \qquad \Longleftrightarrow \qquad \Lambda_- > \Lambda_+ .
\end{equation}
However, this hierarchy of curvature scales, although necessary, is not sufficient. Further stability analysis reveals an additional constraint on the exterior mass parameter,
\begin{equation}\label{mass_bound}
    \frac{m}{L_-}<2\left(1-\frac{L_-^2}{L_+^2}\right).
\end{equation}
In the limit of vanishing exterior cosmological constant, $L_+\to+\infty$, the bound~\eqref{mass_bound} reduces to
\begin{equation}\label{black hole mass upper bound}
    \frac{m}{L_-}<2
\end{equation}
which reproduces the numerical conjecture reported in Ref.~\cite{Uchikata:2012zs}.\\

So far, we have not yet imposed the requirement that the exterior geometry represent a genuine black hole. For a Reissner--Nordström--de Sitter exterior, this requires the existence of three distinct positive horizons, with the shell located inside the inner horizon. After introducing the dimensionless variables
\begin{equation}\label{normalized parameters}
    R \to \frac{R}{L_-}, \quad
    m \to \frac{m}{L_-}, \quad
    Q \to \frac{Q}{L_-}, \quad
    M \to \frac{M}{L_-}, \quad
    L \to \frac{L_+}{L_-},
\end{equation}
the equilibrium conditions $V(R)=0$ and $V'(R)=0$ can be solved analytically for the exterior mass and charge
\begin{align}
    m &= \frac{M(1-2R^2)}{\sqrt{1-R^2}}
    +2R^3\left(1-\frac{1}{L^2}\right), \\
    Q^2 &= M^2-\frac{2MR^3}{\sqrt{1-R^2}}
    +3R^4\left(1-\frac{1}{L^2}\right).
\end{align}
Moreover, the second derivative of the effective potential at equilibrium takes the form
\begin{equation}
    V''(R)= -\frac{2(3-2R^2)}{1-R^2}+\frac{12R\sqrt{1-R^2}}{M} \left(1-\frac{1}{L^2}\right).
\end{equation}

The condition $L>1$ is necessary for radial stability, but the existence of a black-hole exterior imposes a stronger restriction. As shown analytically in Appendix~\ref{app:lower_bound}, stationary and radially stable thin-shell black-hole configurations exist if and only if
\begin{equation}
    L>\sqrt{3}.
\end{equation}
Thus, the lower bound on the exterior curvature radius is obtained analytically rather than numerically. Representative stable configurations spanning different values of the exterior curvature radius are also provided in Table~\ref{tab:stable_configs} of Appendix~\ref{app:lower_bound}.

Figures~\ref{fig: ranges}(a) and~\ref{fig: ranges}(b) display a representative portion of the allowed parameter space in the interval $1.79\leq L\leq 2$. They are included only to illustrate the behavior of the admissible region near the lower bound. The limiting value $L=\sqrt{3}$ corresponds to a degenerate configuration in which the three horizons coalesce and the shell approaches the horizon. This limiting configuration does not describe a static timelike shell.

\begin{figure}[h]
	\centering
	\subfigure[]{\includegraphics[width=0.49\textwidth]{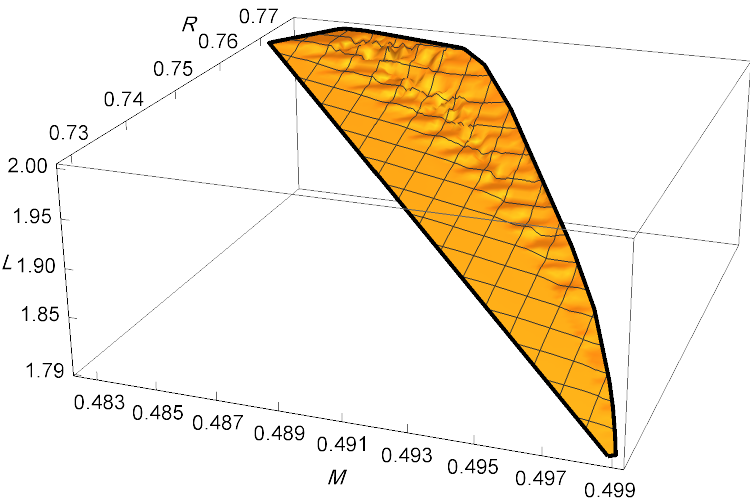}}
	\subfigure[]{\includegraphics[width=0.49\textwidth]{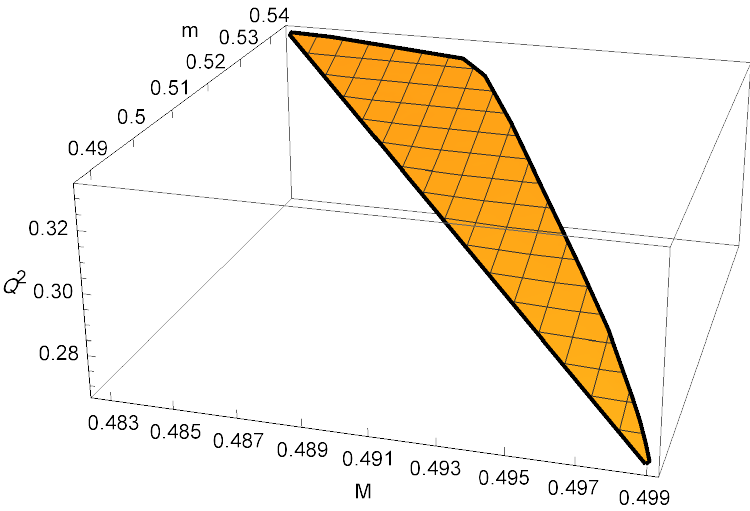}}
	\caption{Representative portions of the admissible parameter space for stationary, radially stable thin-shell black-hole configurations: (a) $(M,R,L)$ and (b) $(M,Q^2,m)$. The plots are shown for the restricted interval $1.79\leq L\leq 2$ in order to visualize the structure of the allowed region near the analytical lower bound $L=\sqrt{3}$.}
    \label{fig: ranges}
\end{figure}
\begin{figure}[h]
	\centering
	\subfigure[]{\includegraphics[width=0.49\textwidth]{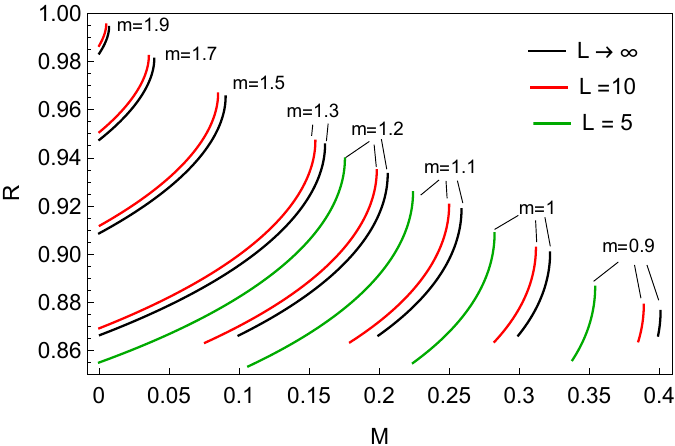}}
	\subfigure[]{\includegraphics[width=0.49\textwidth]{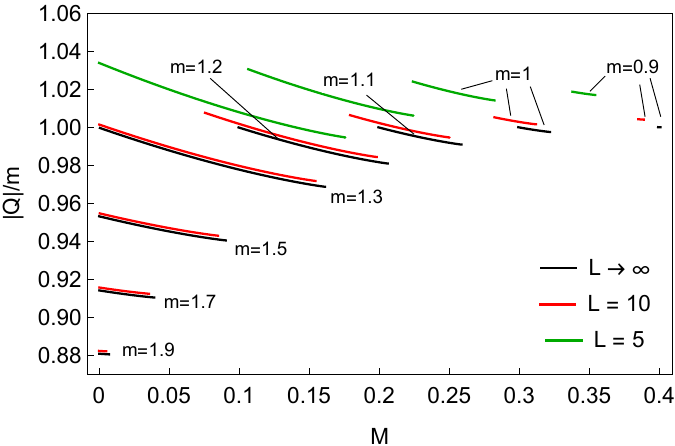}}
	\caption{(a) Normalized shell radius and (b) the ratio $|Q|/m$ as functions of the normalized
shell mass for selected values of $m$ and $L$. These plots illustrate how the family of stationary
configurations changes with the exterior de Sitter scale.}	
    \label{fig: set1}
\end{figure}

The influence of the exterior cosmological constant on the family of stationary solutions is illustrated in Figs.~\ref{fig: set1}(a) and~\ref{fig: set1}(b), which show the dependence of the shell radius and the charge-to-mass ratio on the shell mass for selected values of $m$ and $L$. These plots provide a representative visualization of how the allowed configurations are deformed as the exterior de Sitter scale is varied. In the limit $L\to\infty$, the present model reduces to the regular Reissner--Nordström thin-shell case.

\subsection{Causal structure, vacuum-energy hierarchy,
and mass parameters}

The stable black-hole configurations considered here have the shell located in the inner static region of the exterior Reissner--Nordström--de Sitter geometry, with $R<r_-<r_+<r_c$. The region between the inner and outer black-hole horizons, $r_-<r<r_+$, is nonstatic, with $f_+<0$. Consequently, the shell and the regular de Sitter core are hidden behind the event horizon and cannot communicate with observers in the corresponding exterior static region.

An important consequence is that the regular and singular configurations are indistinguishable through observations restricted to their identical stationary exterior geometries. In particular, exterior observations alone cannot establish whether the central region contains a curvature singularity or a regular core. This illustrates a degeneracy in the inference of black-hole interior structure from external gravitational measurements. The present construction provides an explicit realization of this possibility for a Reissner--Nordström--de Sitter exterior, subject to the existence and radial stability conditions derived above.

A further physical implication concerns the difference between the interior and exterior vacuum
energies. Our stability analysis establishes the necessary condition $\Lambda_->\Lambda_+$, indicating that the effective vacuum energy density of the regular core must exceed that of the exterior region. Thus, the vacuum-energy mismatch is not merely an additional parameter of the construction, but plays a direct role in determining whether a static configuration can be radially stable. Such a hierarchy may be interpreted phenomenologically in terms of distinct effective vacuum states in the high-density interior and the exterior region. The microscopic origin of this transition, however, lies beyond the scope of the present model.

Finally, it is important to distinguish the proper mass $M$ of the shell from the exterior gravitational mass parameter $m$. The former is defined by $M=4\pi R^2\sigma$, whereas the latter characterizes the exterior geometry. For a static shell, the junction equation gives
\begin{equation}
    \sqrt{f_-(R)}-\sqrt{f_+(R)}=\frac{M}{R},
\end{equation}
which can be rearranged to yield
\begin{equation}\label{shell_exterior_mass}
    m=M\sqrt{1-\frac{R^2}{L_-^2}} +\frac{Q^2-M^2}{2R} +\frac{R^3}{2} \left(\frac{1}{L_-^2}-\frac{1}{L_+^2}\right).
\end{equation}
This relation shows that the exterior mass parameter is not simply the proper mass of the shell. It also incorporates the effects of the electromagnetic field, gravitational binding, and the difference between the interior and exterior vacuum energies.

\section{Regular Reissner--Nordström--de Sitter black hole from a modified density profile}
\label{sec:modified_density_profile}

In the previous sections, the regular charged black hole configuration was obtained by matching an interior de Sitter spacetime to an exterior Reissner--Nordström--de Sitter geometry across a timelike thin shell. This construction makes the role of the junction conditions explicit and allows one to analyze the stability of the shell in a direct way. In the present section, we describe a complementary route to a regular Reissner--Nordström--de Sitter black hole. Instead of imposing junction conditions, we use  a modification of the energy-density profile~\cite{bib:yi2026epjc}. The basic idea is to start from the matter distribution supporting the singular Reissner--Nordström--de Sitter geometry and then deform this distribution in such a way that the singular behavior near the center is removed while the standard asymptotic behavior is preserved.

The matter distribution corresponding to the metric~\eqref{fplus} follows from the Einstein equations and takes the form\footnote{In this section, the metric~\eqref{fplus} is assumed to describe the entire spacetime.
}
\begin{eqnarray} \label{eq:density}
\rho&=&\frac{2\mathcal{M}'}{r^2}=\frac{Q^2}{r^4}+\Lambda,\nonumber \\
P&=&-\frac{\mathcal{M}''}{r}=\frac{Q^2}{r^4}-\Lambda,\nonumber \\
\Lambda&\equiv & \frac{3}{L_+^2}.
\end{eqnarray}
Therefore, the singular Reissner--Nordström--de Sitter solution is supported by the equation of state
\begin{equation} \label{eq:eos}
P=\rho-2\Lambda.
\end{equation}

The metric~\eqref{fplus} clearly describes a singular black hole. The singularity originates from the divergent electromagnetic contribution to the energy density as $r\to 0$. We now assume that, in the high-density central region, the matter sector undergoes an effective transition to a de Sitter core. In such a core the equation of state approaches $P=-\rho$, and this behavior prevents the formation of a curvature singularity. Hence, the energy-density profile has to be modified in two simultaneous ways: it must remain finite at the center, while at large distances it must reproduce the standard Reissner--Nordström--de Sitter behavior given in Eq.~\eqref{eq:density}.

For this purpose, we introduce a regularization parameter $\beta$ and deform the energy density as follows:
\begin{equation} \label{eq:den}
\rho(r)\rightarrow \rho(r,\beta)=\frac{Q^2}{(r^3+\beta^3)^\frac{4}{3}}+\Lambda.
\end{equation}

It is evident that the original density profile~\eqref{eq:density} is recovered in the limit $\beta\to 0$. The parameter $\beta$ therefore controls the deviation from the singular Reissner-Nordström-de Sitter distribution and determines the size of the regularized central region.

\begin{figure}[h]
    \centering
    \includegraphics[width=1\linewidth]{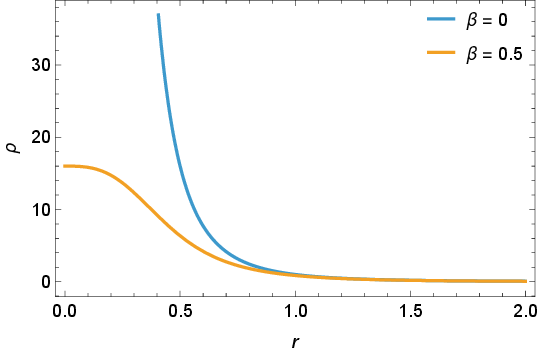}
    \caption{Energy density profiles for the singular Reissner--Nordström--de Sitter solution and for the regularized model. The original density profile diverges as $r\to 0$, whereas the modified profile remains finite at the center. Away from the central region, the two profiles rapidly become almost indistinguishable.}
    \label{fig:densityProfile}
\end{figure}
In a more general construction one could consider a family of profiles of the form
\begin{equation}
\rho=\frac{Q^2}{(r^\alpha+\beta^\alpha)^{\frac{4}{\alpha}}},
\end{equation}
but here we choose $\alpha=3$. This choice is technically convenient and makes the resulting solution closely related to the generalized Hayward-type class~\cite{bib:yi2026epjc}.

Applying the Einstein equation
\begin{equation}
\rho=\frac{2\mathcal{M}'}{r^2}
\end{equation}
to the regularized density profile~\eqref{eq:den}, we obtain the mass function
\begin{equation} \label{eq:mass}
\mathcal{M}(r)=m-\frac{Q^2}{2(r^3+\beta^3)^\frac{1}{3}}+\frac{r^3}{2L_+^2}.
\end{equation}
This expression reduces to the Reissner-Nordström-de Sitter mass function in Eq.~\eqref{fplus} when the regularization is removed, namely in the limit $\beta\to 0$. Thus, the deformation does not change the asymptotic charged de Sitter behavior of the solution, but only modifies the strong-field central region.

The energy density and pressure corresponding to the mass function~\eqref{eq:mass} are given by
\begin{eqnarray} \label{eq:press}
\rho&=&\frac{Q^2}{(r^3+\beta^3)^\frac{4}{3}}+\Lambda, \\
P&=&-\frac{Q^2}{(r^3+\beta^3)^\frac{4}{3}}+\frac{2Q^2r^3}{(r^3+\beta^3)^\frac{7}{3}}-\Lambda.
\end{eqnarray}
It is important to examine the behavior of the matter near the center. Expanding the ratio of the tangential pressure to the energy density at $r\to 0$, we obtain
\begin{equation}
\frac{P}{\rho}+1=
\frac{2Q^2}{Q^2+\Lambda\beta^4}
\left(\frac{r}{\beta}\right)^3
+O\left(\frac{r^6}{\beta^6}\right).
\end{equation}
Therefore, the equation of state approaches $P=-\rho$ at the center, with the leading correction proportional to $r^3$. This confirms the de Sitter behavior of the regular core.

These expressions can be represented in terms of the following equation of state:
\begin{equation}\label{eq:eos1}
P=\rho-2\Lambda
-2\beta^3(Q^2)^{-3/4}(\rho-\Lambda)^{7/4}.
\end{equation}
This is a generalized polytropic-type equation of state. In the limit $\beta\to 0$, the last term vanishes and Eq.~\eqref{eq:eos1} reduces to Eq.~\eqref{eq:eos}, which supports the singular Reissner-Nordström-de Sitter geometry. Therefore, the regular solution can be understood as being generated by a nonlinear deformation of the matter equation of state in the high-density regime.

It is important to emphasize that the mass function~\eqref{eq:mass} does not describe a regular black hole for an arbitrary value of the parameter $\beta$. Although the energy density is finite at the origin, the regularity of the geometry requires the curvature invariants to remain finite as well. In particular, the Kretschmann scalar is finite at $r\to 0$ only if the mass function satisfies the additional condition
\begin{equation}
\lim\limits_{r\to 0}\mathcal{M}(r)=0.
\end{equation}
Using Eq.~\eqref{eq:mass}, this condition fixes the regularization parameter as
\begin{equation}
\beta=\frac{Q^2}{2m}.
\end{equation}
Thus, regularity is not achieved by an arbitrary smoothing of the density profile; instead, it imposes a definite relation between the charge, the total mass, and the regularization scale.

Consequently, we arrive at a spacetime which behaves at large distances as the Reissner-Nordström-de Sitter solution, but possesses a de Sitter core at the center and therefore describes a regular charged black hole:
\begin{eqnarray}
ds^2&=&-f(r)dt^2+f^{-1}(r)dr^2+r^2d\Omega^2,\nonumber \\
f(r)&=&1-\frac{2m}{r}
+\frac{Q^2}{r\left(r^3+\frac{Q^6}{8m^3}\right)^{1/3}}
-\frac{r^2}{L_+^2}.
\end{eqnarray}It is important to emphasize that the two constructions considered in this paper represent different regularization mechanisms for the same singular Reissner--Nordstrom--de Sitter solution, rather than two descriptions of the same spacetime. In the thin-shell approach, the singular region is replaced by a de Sitter core with an independent cosmological constant $\Lambda_-$, which is matched to the exterior geometry characterized by $\Lambda_+$. The stability of this configuration requires $\Lambda_->\Lambda_+$. In contrast, the modified-density approach does not introduce a junction surface or an independent interior cosmological constant. Instead, the de Sitter core emerges from a smooth modification of the energy-density profile, with its central density determined by the charge, mass, and exterior cosmological constant. Furthermore, the regularization parameter is not independent, since the regularity condition fixes $\beta=Q^2/(2m)$.

Thus, although both approaches replace the central singularity with a regular de Sitter-like region, they differ in their matter distributions, core structures, and parameter freedom. In particular, the thin-shell model contains an additional junction surface, whereas the smooth model modifies the matter distribution throughout the spacetime. Consequently, the two constructions cannot be directly identified through a simple correspondence between their parameters.

\section{Conclusion}

In this work we investigated two complementary mechanisms for regularizing the Reissner--Nordström--de Sitter black hole. The first construction, based on the Israel junction condition formalism, replaces the singular central region with an interior de Sitter core  separated from the exterior Reissner--Nordström--de Sitter geometry by a pressureless timelike shell.

For this thin-shell configuration, we derived the shell equation of motion and expressed the dynamics in terms of an effective potential. The conditions for static equilibrium and linear radial stability lead to a necessary constraint on the vacuum curvature scales, $L_+>L_-$, or equivalently $\Lambda_->\Lambda_+$. Thus, the mismatch between the interior and exterior cosmological constants is not merely an additional parameter of the model, but a necessary condition for the radial stability of the regular black-hole configuration.

The stability analysis also gives an upper bound on the exterior mass parameter, $\frac{m}{L_-}<2\left(1-\frac{L_-^2}{L_+^2}\right).$ In the asymptotically flat limit $L_+\to\infty$, this reduces to $m/L_-<2$, which is the corresponding bound for the thin shell regular Reissner--Nordström case. The remaining parameter bounds in the Reissner--Nordström limit can be obtained analytically by specializing the Kiselev thin shell analysis of~\cite{Saadati:2020dje} to the appropriate limit, where they reproduce the numerical ranges reported in~\cite{Uchikata:2012zs}.

We also analyzed the horizon structure to identify which stationary shell configurations correspond
to genuine black holes. Requiring three distinct positive exterior horizons, with the shell located
inside the inner horizon, we demonstrated analytically that radially stable thin-shell black-hole configurations exist if and only if $L_+/L_->\sqrt{3}$, or equivalently $\Lambda_->3\Lambda_+$. This exact existence bound is stronger than the condition obtained from radial stability alone and demonstrates that the exterior cosmological constant significantly restricts the admissible parameter space of regular charged black holes.

An important implication of the thin-shell construction is that the existence of a regular interior cannot be established or excluded solely through observations of the stationary exterior geometry. Since the shell and the de Sitter core lie behind the event horizon, the exterior spacetime is identical to that of a conventional singular Reissner--Nordström--de Sitter black hole
with the same exterior parameters. Nevertheless, the internal structure is subject to nontrivial physical constraints. In particular, the vacuum-energy hierarchy $\Lambda_->3\Lambda_+$ provides an exact condition for the existence of stationary, radially stable black-hole configurations of the type considered here. Our analysis therefore constrains possible nonsingular black-hole interiors without implying a directly observable signature of the regular core.

In addition to the thin-shell construction, we have also considered a complementary regularization mechanism based on a smooth modification of the energy-density profile. Starting from the matter distribution corresponding to the singular Reissner--Nordstrom--de Sitter geometry, we introduced a regularized density which remains finite at the center and reproduces the standard Reissner--Nordstrom--de Sitter behavior at large radii. Reconstructing the mass function from the Einstein equations, we obtained a nonsingular geometry with a de Sitter core, while the original singular solution is recovered when the regularization parameter is removed.
This approach shows that the formation of a regular core can be described not only by matching two geometries across a thin shell, but also by an effective high-density deformation of the matter source itself. In this case, the regularization is encoded in a generalized polytropic-type equation of state, which reduces to the usual Reissner--Nordstrom--de Sitter equation of state in the singular limit. Thus, the modified-density construction provides a smooth counterpart to the thin-shell picture and supports the same physical interpretation: the central singularity is replaced by a vacuum-like de Sitter phase generated by matter effects in the strong-field region.

\appendix
\section{Analytical lower bound on the exterior
curvature radius}
\label{app:lower_bound}

In this appendix, we derive the exact lower bound on the exterior curvature radius for the existence of stationary, radially stable thin-shell black-hole configurations. We employ the dimensionless variables introduced in Sec.~\ref{sec:Regular Reissner--Nordström--de Sitter Configuration}, setting $L_-=1$ and $L=L_+/L_-$. The exterior metric function is
\begin{equation}
    f_+(r)=1-\frac{2m}{r}+\frac{Q^2}{r^2}-\frac{r^2}{L^2}.
\end{equation}

The horizon radii are determined by the positive roots of the quartic polynomial
\begin{equation}\label{app:horizon_poly}
    P(r)=r^4-L^2r^2+2mL^2r-Q^2L^2=0.
\end{equation}
For the black-hole configurations considered here, the exterior must possess three distinct positive horizons, $r_-<r_+<r_c$, with the shell located inside the inner horizon, $R<r_-$.

Since the coefficient of $r^3$ in Eq.~\eqref{app:horizon_poly} vanishes, the fourth root is $-(r_-+r_++r_c)$. Using Vieta's relations for the coefficient of $r^2$, we obtain
\begin{equation}
    L^2=r_-^2+r_+^2+r_c^2+r_-r_++r_-r_c+r_+r_c.
\end{equation}
Because $0<r_-<r_+<r_c$, each term on the right-hand side is at least $r_-^2$, with strict inequality for five of the six terms. Together with $R<r_-$, this yields
\begin{equation}\label{app:horizon_bound}
    L^2>6r_-^2>6R^2.
\end{equation}

To incorporate the equilibrium conditions, we introduce the dimensionless variables \begin{equation}
    y=R^2,\qquad z=L^2,\qquad x=\frac{M}{R\sqrt{1-R^2}}.
\end{equation}
The interior staticity condition requires $0<y<1$, while positivity of the shell mass and the static junction condition imply $0<x<1$.

The equilibrium conditions $V(R)=V'(R)=0$ allow us to express the exterior mass and charge in terms of these variables:
\begin{align}
    m&=R\left[x(1-2y)+2y\left(1-\frac{1}{z}\right)\right],\\
    Q^2&=R^2\left[(1-y)x^2-2xy+3y\left(1-\frac{1}{z}\right)\right].
    \label{app:equilibrium_parameters}
\end{align}

\subsection{Necessity of the lower bound}

We first demonstrate that the existence of three distinct positive horizons outside a stationary timelike shell requires $z>3$. Let $\delta=1-x$, with $0<\delta<1$, and introduce an auxiliary dimensionless radial variable $\xi$ through
\begin{equation}
    r=R(1+\delta\xi).
\end{equation}
Since $\delta>0$, the condition $r>R$ is equivalent to $\xi>0$. Therefore, the existence of three distinct horizons outside the shell requires three distinct positive roots in $\xi$. Substituting the equilibrium expressions for $m$ and $Q^2$ into Eq.~\eqref{app:horizon_poly}, we obtain
\begin{align}
    F(\xi)&\equiv\frac{P\bigl(R(1+\delta\xi)\bigr)}{R^2\delta^2} \nonumber\\
    &=\delta^2y\xi^4+4\delta y\xi^3+(6y-z)\xi^2 \nonumber\\
    &\quad-2z(1-2y)\xi-z(1-y).
    \label{app:transformed_poly}
\end{align}

Suppose that $z\leq3$. From Eq.~\eqref{app:horizon_bound}, we have $y<z/6$, and consequently \begin{equation}
    y<\frac{z}{6}\leq\frac12.
\end{equation}
It follows that
\begin{equation}
    6y-z<0,\qquad 1-2y>0.
\end{equation}
Hence, the coefficients of Eq.~\eqref{app:transformed_poly}, ordered by descending powers of $\xi$, have the signs
\begin{equation}
    (+,+,-,-,-).
\end{equation}
Since there is only one sign change, Descartes' rule of signs implies that $F(\xi)$ has at most one positive real root, counted with multiplicity. This contradicts the requirement of three distinct positive horizons outside the shell. We therefore conclude that $z>3$ is necessary, or equivalently,
\begin{equation}\label{app:necessary_bound}
    L>\sqrt{3}.
\end{equation}

Notice that this necessary condition follow from the equilibrium equations and the horizon structure alone, without imposing radial stability.

\subsection{Existence of stable configurations}

We now show that the condition \eqref{app:necessary_bound} is also sufficient for the existence of at least one stationary, radially stable thin-shell configuration with three distinct positive exterior horizons. To establish sufficiency, it is not necessary to characterize the entire admissible parameter space. Instead, we construct a family of equilibrium configurations that contains an admissible solution for every $z>3$. The equilibrium expression for the second derivative of the effective potential is \begin{equation}\label{app:Vsecond}
    V''(R)=-\frac{2(3-2y)}{1-y}+\frac{12}{x}\left(1-\frac{1}{z}\right).
\end{equation}
We begin by considering the limiting value $x=1$, corresponding to the boundary $f_+(R)=0$. Requiring marginal radial stability, $V''(R)=0$, at this boundary gives
\begin{equation}
    \frac{3-2y}{1-y}=6\left(1-\frac{1}{z}\right),
\end{equation}
whose solution is
\begin{equation}
    y=\frac{3z-6}{4z-6}.
\end{equation}

This choice also makes the transformed horizon polynomial develop a positive double root in the limiting case $\delta\to0^+$. We therefore move slightly away from this boundary by taking $x=1-\delta$, with a sufficiently small positive $\delta$. As shown below, this construction simultaneously ensures radial stability and the existence of three distinct exterior horizons. Thus, for any fixed $z>3$, we consider the family
\begin{equation}\label{app:family}
    y=\frac{3z-6}{4z-6},\qquad x=1-\delta,\qquad 0<\delta\ll1.
\end{equation}
For $z>3$, this choice satisfies $0<y<1$ and $y<z/6$.

The constructed family also satisfies the positivity requirements for the exterior mass and squared charge. Indeed, in the limit $\delta\to0^+$, the equilibrium relations yield \begin{equation}
    \frac{m}{R}=1-\frac{2y}{z}>\frac{2}{3},\qquad\frac{Q^2}{R^2}=1-\frac{3y}{z}>\frac{1}{2},
\end{equation}
where we have used $y<z/6$. By continuity, both quantities remain positive for sufficiently small positive $\delta$. Substituting Eq.~\eqref{app:family} into Eq.~\eqref{app:Vsecond}, we obtain \begin{equation}\label{app:stability}
    V''(R)=12\left(1-\frac{1}{z}\right)\frac{\delta}{1-\delta}>0.
\end{equation}
Thus, the shell is radially stable for every sufficiently small positive $\delta$. Furthermore, using the equilibrium relations, the metric functions at the shell satisfy
\begin{equation}
    f_-(R)=1-y>0,\qquad f_+(R)=(1-y)\delta^2>0.
\end{equation}
These expressions also verify the original unsquared junction condition:
\begin{equation}
    \sqrt{f_-(R)}-\sqrt{f_+(R)} =(1-\delta)\sqrt{1-y}=\frac{M}{R}.
\end{equation}
The shell therefore remains timelike and satisfies the required junction condition.

The limit $\delta\to0^+$ is used only to construct the admissible family. At $\delta=0$, the shell coincides with a horizon and no longer represents a static timelike hypersurface. The admissible
configurations consequently require $\delta>0$.

It remains to verify the existence of three distinct positive horizons outside the shell. For the family~\eqref{app:family}, the polynomial~\eqref{app:transformed_poly} can be rewritten as \begin{equation}\label{app:factorized_poly}
    F(\xi)=(6y-z)(\xi-\xi_0)^2+4\delta y\xi^3+\delta^2y\xi^4,
\end{equation}
where
\begin{equation}
    \xi_0=\frac{z}{2(z-3)}>0.
\end{equation}

Since $6y-z<0$, we have $F(0)<0$. Moreover,
\begin{equation}
    F(\xi_0)=4\delta y\xi_0^3+\delta^2y\xi_0^4>0,
\end{equation}
whereas
\begin{equation}
    \lim_{\delta\to0^+}F(2\xi_0)=(6y-z)\xi_0^2=-z(1-y)<0.
\end{equation}
Therefore, for sufficiently small positive $\delta$, the following sign relations hold
\begin{equation}
    \begin{aligned}
    F(0)&<0,\\
    F(\xi_0)&>0,\\
    F(2\xi_0)&<0,\\
    \lim_{\xi\to\infty}F(\xi)&=+\infty.
    \end{aligned}
\end{equation}

By continuity, $F(\xi)$ must have at least one positive root in each of the intervals $(0,\xi_0)$, $(\xi_0,2\xi_0)$, and $(2\xi_0,\infty)$. These roots are distinct and correspond to three positive horizon radii satisfying
\begin{equation}
    R<r_-<r_+<r_c.
\end{equation}
Since the quartic horizon polynomial also has a negative root, these are precisely its three positive roots. Hence, for every $z>3$, a sufficiently small positive $\delta$ yields a stationary, radially stable timelike shell with the required black-hole horizon structure. Combining the necessity and existence results, we conclude that the exact condition for the existence of stationary, radially stable thin-shell black-hole configurations with three distinct positive exterior horizons and $R<r_-$ is
\begin{equation}
    \frac{L_+}{L_-}>\sqrt{3}.
\end{equation}
Equivalently, in terms of the interior and exterior effective cosmological constants,
\begin{equation}
    \Lambda_->3\Lambda_+.
\end{equation}

\subsection{Representative stable configurations}

To complement the analytical existence result, we present several representative stationary, radially stable black-hole configurations in Table~\ref{tab:stable_configs}. The parameters are obtained from the equilibrium relations using the family constructed above, with $x=1-\delta$ and
$y=(3L^2-6)/(4L^2-6)$.

For each configuration, the exterior horizon equation has three distinct positive roots, satisfying $R<r_-<r_+<r_c$, and the second derivative of the effective potential is positive. These examples illustrate the admissible parameter space and are not used to establish the analytical lower bound.

\begin{table}[h]
\centering
\caption{Representative stationary, radially stable thin-shell black-hole configurations in units
$L_-=1$. All numerical values are rounded. The parameter $\delta$ specifies the exact generating family and allows the configurations to be reconstructed from the analytical equilibrium relations.}
\label{tab:stable_configs}
\resizebox{\textwidth}{!}{%
\begin{tabular}{cccccccccc}
\hline
$L$ & $\delta$ & $R$ & $M$ & $m$ &
$|Q|$ & $V''(R)$ & $r_-$ & $r_+$ & $r_c$ \\
\hline
1.75 & $4\times10^{-7}$
& 0.714142843 & 0.499899790 & 0.476289559
& 0.505181325 & $3.23\times10^{-6}$
& 0.714149288 & 0.714150567 & 0.715003584 \\

1.80 & $9\times10^{-5}$
& 0.731083277 & 0.498764629 & 0.489883243
& 0.519594507 & $7.47\times10^{-4}$
& 0.731458965 & 0.731656372 & 0.741414411 \\

2 & $3\times10^{-3}$
& 0.774596669 & 0.488428255 & 0.542682426
& 0.575084481 & 0.027081
& 0.778551588 & 0.780511512 & 0.889198555 \\

3 & 0.035
& 0.836660027 & 0.442218555 & 0.718226152
& 0.746005306 & 0.386874
& 0.856314773 & 0.862004057 & 1.883773706 \\

5 & 0.060
& 0.856762834 & 0.415331193 & 0.830513079
& 0.843396982 & 0.735319
& 0.884117761 & 0.888261508 & 3.954195844 \\

10 & 0.080
& 0.863824573 & 0.400373450 & 0.884959773
& 0.888530941 & 1.033043
& 0.898226607 & 0.900806010 & 9.019240627 \\
\hline
\end{tabular}%
}
\end{table}

\acknowledgments

We acknowledge the support of the National Natural Science Foundation of China under Grant No.  12075059, as well as the start-up fund of USTC. 



 \bibliographystyle{JHEP}
 \bibliography{biblio.bib}

@article{bib:eht1,
    author = "Akiyama, Kazunori and others",
    collaboration = "Event Horizon Telescope",
    title = "{First M87 Event Horizon Telescope Results. I. The Shadow of the Supermassive Black Hole}",
    eprint = "1906.11238",
    archivePrefix = "arXiv",
    primaryClass = "astro-ph.GA",
    doi = "10.3847/2041-8213/ab0ec7",
    journal = "Astrophys. J. Lett.",
    volume = "875",
    pages = "L1",
    year = "2019"
}

@article{bib:eht2,
    author = "Akiyama, Kazunori and others",
    collaboration = "Event Horizon Telescope",
    title = "{First Sagittarius A* Event Horizon Telescope Results. I. The Shadow of the Supermassive Black Hole in the Center of the Milky Way}",
    eprint = "2311.08680",
    archivePrefix = "arXiv",
    primaryClass = "astro-ph.HE",
    doi = "10.3847/2041-8213/ac6674",
    journal = "Astrophys. J. Lett.",
    volume = "930",
    number = "2",
    pages = "L12",
    year = "2022"
}

@article{bib:penrose,
    author = {Penrose, Roger},
    title = {Gravitational collapse and space-time singularities},
    doi = {10.1103/PhysRevLett.14.57},
    journal = {Phys. Rev. Lett.},
    volume = {14},
    pages = {57--59},
    year = {1965}
}

@book{bib:joshi,
    editor = {Joshi, Pankaj S.},
    title = {Gravitational Collapse and Spacetime Singularities},
    doi = {10.1017/CBO9780511536274},
    isbn = {978-1-107-40536-3, 978-0-521-87104-4, 978-0-511-37283-4},
    publisher = {Cambridge University Press},
    series = {Cambridge Monographs on Mathematical Physics},
    month = {9},
    year = {2012}
}

@article{bib:joshi_review,
    author = {Joshi, Pankaj S. and Malafarina, Daniele},
    title = {Recent developments in gravitational collapse and spacetime singularities},
    eprint = {1201.3660},
    archivePrefix = {arXiv},
    primaryClass = {gr-qc},
    doi = {10.1142/S0218271811020792},
    journal = {Int. J. Mod. Phys. D},
    volume = {20},
    pages = {2641--2729},
    year = {2011}}

@article{bib:dei2022epjc,
    author = {Dey, Dipanjan and Joshi, Pankaj S. and Mosani, Karim and Vertogradov, Vitalii},
    title = {Causal structure of singularity in non-spherical gravitational collapse},
    eprint = {2103.07190},
    archivePrefix = {arXiv},
    primaryClass = {gr-qc},
    doi = {10.1140/epjc/s10052-022-10401-1},
    journal = {Eur. Phys. J. C},
    volume = {82},
    number = {5},
    pages = {431},
    year = {2022}
}

@article{bib:inhomo,
    author = {Singh, T. P. and Joshi, P. S.},
    title = {The Final fate of spherical inhomogeneous dust collapse},
    eprint = {gr-qc/9409062},
    archivePrefix = {arXiv},
    doi = {10.1088/0264-9381/13/3/019},
    journal = {Class. Quant. Grav.},
    volume = {13},
    pages = {559--572},
    year = {1996}
}

@article{bib:inhomo2,
    author = {Jhingan, Sanjay and Joshi, P. S. and Singh, T. P.},
    title = {The Final fate of spherical inhomogeneous dust collapse. 2. Initial data and causal structure of singularity},
    eprint = {gr-qc/9604046},
    archivePrefix = {arXiv},
    doi = {10.1088/0264-9381/13/11/019},
    journal = {Class. Quant. Grav.},
    volume = {13},
    pages = {3057--3068},
    year = {1996}
}

@INPROCEEDINGS{bib:bardeen,
       author = {Bardeen, James},
        title = {Non-singular general relativistic gravitational collapse},
    booktitle = {Proceedings of the 5th International Conference on Gravitation and the Theory of Relativity},
         year = {1968},
        month = {sep},
        pages = {87},
       url = {https://ui.adsabs.harvard.edu/abs/1968qtr..conf...87B}
}

@article{bib:garcia,
    author = {Ayon-Beato, Eloy and Garcia, Alberto},
    title = {The Bardeen model as a nonlinear magnetic monopole},
    eprint = {gr-qc/0009077},
    archivePrefix = {arXiv},
    doi = {10.1016/S0370-2693(00)01125-4},
    journal = {Phys. Lett. B},
    volume = {493},
    pages = {149--152},
    year = {2000}
}

@article{bib:bronnikov,
    author = {Bronnikov, Kirill A.},
    title = {Regular magnetic black holes and monopoles from nonlinear electrodynamics},
    eprint = {gr-qc/0006014},
    archivePrefix = {arXiv},
    doi = {10.1103/PhysRevD.63.044005},
    journal = {Phys. Rev. D},
    volume = {63},
    pages = {044005},
    year = {2001}
}

@article{bib:angel2025podu,
    author = "Vertogradov, Vitalii and Rincon, Angel",
    title = "{Energy extraction and evolution of regular black holes: The case of Bardeen spacetime}",
    eprint = "2508.14489",
    archivePrefix = "arXiv",
    primaryClass = "gr-qc",
    doi = "10.1016/j.dark.2025.102066",
    journal = "Phys. Dark Univ.",
    volume = "50",
    pages = "102066",
    year = "2025"
}

@ARTICLE{bib:gliner,
       author = {Gliner, E.~B.},
        title = {Algebraic Properties of the Energy-momentum Tensor and Vacuum-like States o$^{+}$ Matter},
      journal = {Soviet Journal of Experimental and Theoretical Physics},
         year = {1966},
        month = {feb},
       volume = {22},
        pages = {378},
       url = {https://ui.adsabs.harvard.edu/abs/1966JETP...22..378G}
}

@ARTICLE{bib:sakharov,
       author = {Sakharov, A.~D.},
        title = {The Initial Stage of an Expanding Universe and the Appearance of a Nonuniform Distribution of Matter},
      journal = {Soviet Journal of Experimental and Theoretical Physics},
         year = {1966},
        month = {jan},
       volume = {22},
        pages = {241},
       url = {https://ui.adsabs.harvard.edu/abs/1966JETP...22..241S}
}

@article{bib:dym,
    author = {Dymnikova, I.},
    title = {Vacuum nonsingular black hole},
    doi = {10.1007/BF00760226},
    journal = {Gen. Rel. Grav.},
    volume = {24},
    pages = {235--242},
    year = {1992}
}

@article{bib:dym2,
    author = {Dymnikova, Irina},
    title = {Cosmological term as a source of mass},
    eprint = {gr-qc/0112052},
    doi = {10.1088/0264-9381/19/4/306},
    journal = {Class. Quant. Grav.},
    volume = {19},
    pages = {725--740},
    year = {2002}
}

@article{bib:hay,
    author = {Hayward, Sean A.},
    title = {Formation and evaporation of regular black holes},
    eprint = {gr-qc/0506126},
    archivePrefix = {arXiv},
    doi = {10.1103/PhysRevLett.96.031103},
    journal = {Phys. Rev. Lett.},
    volume = {96},
    pages = {031103},
    year = {2006}
}

@article{bib:khlopov1,
    author = {Konoplich, R. V. and Rubin, S. G. and Sakharov, A. S. and Khlopov, M. Yu.},
    title = {Formation of black holes in first-order phase transitions as a cosmological test of symmetry-breaking mechanisms},
    journal = {Phys. Atom. Nucl.},
    volume = {62},
    pages = {1593--1600},
    year = {1999}
}

@article{bib:khlopov2,
    author = {Khlopov, M. Yu. and Konoplich, R. V. and Rubin, S. G. and Sakharov, A. S.},
    editor = {Melnikov, V. N.},
    title = {First order phase transitions as a source of black holes in the early universe},
    eprint = {hep-ph/9912422},
    archivePrefix = {arXiv},
    journal = {Grav. Cosmol.},
    volume = {2},
    pages = {S1},
    year = {1999}
}

@article{bib:khlopov3,
    author = {Khlopov, Maxim Yu.},
    title = {Primordial Black Holes},
    eprint = {0801.0116},
    archivePrefix = {arXiv},
    primaryClass = {astro-ph},
    doi = {10.1088/1674-4527/10/6/001},
    journal = {Res. Astron. Astrophys.},
    volume = {10},
    pages = {495--528},
    year = {2010}
}

@article{bib:khlopov4,
    author = {Belotsky, K. M. and Dmitriev, A. D. and Esipova, E. A. and Gani, V. A. and Grobov, A. V. and Khlopov, M. Yu. and Kirillov, A. A. and Rubin, S. G. and Svadkovsky, I. V.},
    title = {Signatures of primordial black hole dark matter},
    eprint = {1410.0203},
    archivePrefix = {arXiv},
    primaryClass = {astro-ph.CO},
    doi = {10.1142/S0217732314400057},
    journal = {Mod. Phys. Lett. A},
    volume = {29},
    number = {37},
    pages = {1440005},
    year = {2014}
}

@article{bib:khlopov5,
    author = {Dymnikova, Irina and Khlopov, Maxim},
    title = {Regular black hole remnants and graviatoms with de Sitter interior as heavy dark matter candidates probing inhomogeneity of early universe},
    eprint = {1510.01351},
    archivePrefix = {arXiv},
    primaryClass = {gr-qc},
    doi = {10.1142/S0218271815450029},
    journal = {Int. J. Mod. Phys. D},
    volume = {24},
    number = {13},
    pages = {1545002},
    year = {2015}
}

@article{bib:ali2025cqg,
    author = {Vertogradov, Vitalii and \"Ovg\"un, Ali},
    title = {Exact regular black hole solutions with de Sitter cores and Hagedorn fluid},
    eprint = {2408.02699},
    archivePrefix = {arXiv},
    primaryClass = {gr-qc},
    doi = {10.1088/1361-6382/ada082},
    journal = {Class. Quant. Grav.},
    volume = {42},
    number = {2},
    pages = {025024},
    year = {2025}
}

@inproceedings{bib:ansoldy,
    author = {Ansoldi, Stefano},
    title = {Spherical black holes with regular center: A Review of existing models including a recent realization with Gaussian sources},
    booktitle = {Conference on Black Holes and Naked Singularities},
    eprint = {0802.0330},
    archivePrefix = {arXiv},
    primaryClass = {gr-qc},
    reportNumber = {KUNS-2108},
    month = {2},
    year = {2008}
}

@article{bib:regular,
    author = {Fan, Zhong-Ying and Wang, Xiaobao},
    title = {Construction of Regular Black Holes in General Relativity},
    eprint = {1610.02636},
    archivePrefix = {arXiv},
    primaryClass = {gr-qc},
    doi = {10.1103/PhysRevD.94.124027},
    journal = {Phys. Rev. D},
    volume = {94},
    number = {12},
    pages = {124027},
    year = {2016}
}

@article{bib:zaslav,
    author = {Zaslavskii, O. B.},
    title = {Regular black holes and energy conditions},
    doi = {10.1016/j.physletb.2010.04.031},
    journal = {Phys. Lett. B},
    volume = {688},
    pages = {278--280},
    year = {2010},
    eprint = {1004.2362},
    archivePrefix = {arXiv},
    primaryClass = {gr-qc}
}

@article{bib:vertogradov2025podu,
    author = "Vertogradov, Vitalii",
    title = "{Regular Black Hole from gravitational collapse of dust and radiation}",
    eprint = "2501.13739",
    archivePrefix = "arXiv",
    primaryClass = "gr-qc",
    doi = "10.1016/j.dark.2025.101881",
    journal = "Phys. Dark Univ.",
    volume = "48",
    pages = "101881",
    year = "2025"
}

@article{bib:ali2025jcap,
    author = {Vertogradov, Vitalii and {\"O}vg{\"u}n, Ali},
    title = "{Regular black hole models in the transition from baryonic matter to quark matter}",
    eprint = "2504.07561",
    archivePrefix = "arXiv",
    primaryClass = "gr-qc",
    doi = "10.1088/1475-7516/2025/06/051",
    journal = "JCAP",
    volume = "06",
    pages = "051",
    year = "2025"
}

@article{bib:vertogradov2025epjc,
    author = "Vertogradov, Vitalii",
    title = "{Gravitational collapse and formation of regular black holes: Dymnikova, Hayward, and beyond}",
    eprint = "2504.19292",
    archivePrefix = "arXiv",
    primaryClass = "gr-qc",
    doi = "10.1140/epjc/s10052-025-14583-2",
    journal = "Eur. Phys. J. C",
    volume = "85",
    number = "8",
    pages = "839",
    year = "2025"
}

@article{Barrabes:1991ng,
    author = "Barrabes, C. and Israel, W.",
    title = "{Thin shells in general relativity and cosmology: The Lightlike limit}",
    doi = "10.1103/PhysRevD.43.1129",
    journal = "Phys. Rev. D",
    volume = "43",
    pages = "1129--1142",
    year = "1991"
}

@article{Israel:1966rt,
    author = "Israel, W.",
    title = "{Singular hypersurfaces and thin shells in general relativity}",
    doi = "10.1007/BF02710419",
    journal = "Nuovo Cim. B",
    volume = "44S10",
    pages = "1",
    year = "1966",
    note = "[Erratum: Nuovo Cim.B 48, 463 (1967)]"
}

@article{Mazur:2004fk,
    author = "Mazur, Pawel O. and Mottola, Emil",
    title = "{Gravitational vacuum condensate stars}",
    eprint = "gr-qc/0407075",
    archivePrefix = "arXiv",
    doi = "10.1073/pnas.0402717101",
    journal = "Proc. Nat. Acad. Sci.",
    volume = "101",
    pages = "9545--9550",
    year = "2004"
}

@article{Mottola:2023jxl,
    author = "Mottola, Emil",
    title = "{Gravitational Vacuum Condensate Stars}",
    eprint = "2302.09690",
    archivePrefix = "arXiv",
    primaryClass = "gr-qc",
    doi = "10.1007/978-981-99-1596-5_8",
    month = "2",
    year = "2023"
}

@article{Visser:2003ge,
    author = "Visser, Matt and Wiltshire, David L.",
    title = "{Stable gravastars: An Alternative to black holes?}",
    eprint = "gr-qc/0310107",
    archivePrefix = "arXiv",
    doi = "10.1088/0264-9381/21/4/027",
    journal = "Class. Quant. Grav.",
    volume = "21",
    pages = "1135--1152",
    year = "2004"
}

@article{MartinMoruno:2011rm,
    author = "Martin Moruno, Prado and Montelongo Garcia, Nadiezhda and Lobo, Francisco S. N. and Visser, Matt",
    title = "{Generic thin-shell gravastars}",
    eprint = "1112.5253",
    archivePrefix = "arXiv",
    primaryClass = "gr-qc",
    doi = "10.1088/1475-7516/2012/03/034",
    journal = "JCAP",
    volume = "03",
    pages = "034",
    year = "2012"
}

@article{Lemos:2011dq,
    author = "Lemos, Jose P. S. and Zanchin, Vilson T.",
    title = {{Regular black holes: Electrically charged solutions, Reissner-Nordstr{\"o}m outside a de Sitter core}},
    eprint = "1104.4790",
    archivePrefix = "arXiv",
    primaryClass = "gr-qc",
    doi = "10.1103/PhysRevD.83.124005",
    journal = "Phys. Rev. D",
    volume = "83",
    pages = "124005",
    year = "2011"
}

@article{Uchikata:2012zs,
    author = "Uchikata, Nami and Yoshida, Shijun and Futamase, Toshifumi",
    title = "{New solutions of charged regular black holes and their stability}",
    eprint = "1209.3567",
    archivePrefix = "arXiv",
    primaryClass = "gr-qc",
    doi = "10.1103/PhysRevD.86.084025",
    journal = "Phys. Rev. D",
    volume = "86",
    pages = "084025",
    year = "2012"
}

@article{Saadati:2020dje,
    author = "Saadati, R. and Shojai, F.",
    title = "{Thin shell collapse in Kiselev geometry}",
    eprint = "2009.06990",
    archivePrefix = "arXiv",
    primaryClass = "gr-qc",
    doi = "10.1088/1361-6382/abfed5",
    journal = "Class. Quant. Grav.",
    volume = "38",
    number = "13",
    pages = "135025",
    year = "2021"
}

@article{bib:yi2026epjc,
    author = "Kudryavcev, Dmitriy and Ling, Yi and Vertogradov, Vitalii",
    title = "{Exact Dynamical Regular Black Holes from Generalized Polytropic Matter}",
    eprint = "2602.10768",
    archivePrefix = "arXiv",
    primaryClass = "gr-qc",
    month = "2",
    year = "2026"
}

\end{document}